\documentstyle[11pt,paspconf,psfig]{article}

\begin{document}

\title{Gas Dynamics in the Galactic Bar Region}
\author{R. Fux}
\affil{Geneva Observatory, CH-1290 Sauverny, Switzerland}
\vspace*{1.4cm}

Many early type barred spiral galaxies exhibit strong dustlanes at the
leading edges of their bar (e.g. NGC 1300, NGC 1433), which from gas
velocity field measurements and numerical hydrodynamical simulations
are known to be the loci of shearing shocks with velocity gaps up to
200~km\,s$^{-1}$ across. As an SBbc type galaxy, the Milky Way is
expected to present similar dustlanes. Self-consistent high resolution
3D $N$-body and SPH simulations allow us to reliably identify their gaseous
traces in the Galactic CO and HI $\ell-V$ distributions. The near-side
branch of these dustlanes corresponds to the connecting arm feature
and lies below ($b<0$) the Galactic plane, while the far side branch,
seen nearly end-on, traces a velocity-elongated feature near
$\ell=-4^{\circ}$ and is located above the plane. The 3-kpc and
135-km\,s$^{-1}$ arms are lateral arms, i.e. inner prolongations of
disc spiral arms passing round the bar and finally intersecting the
dustlanes at their respective latitudes (Figs. 1a-b).
\par Considering that the latter intersections occur close to the
major axis of the bar and within the bar ends, their spatial location
can provide new constraints on the inclination angle $\varphi$ 
and corotation radius $R_{\rm CR}$ of the Galactic bar. Geometrically,
if $\ell_1$ and $\ell_2$ are the longitudes where the 3-kpc and the
135-km\,s$^{-1}$ arms respectively intersect the dustlanes, and $s_1$
and $s_2$ the galactocentric distances of these intersections
(Fig.~1c), then:
\begin{displaymath}
\frac{s_i}{R_{\circ}}=\frac{\sin{\ell_i}}{\sin{(\varphi+\ell_i)}}
\hspace{.5cm}(i=1,\,2),
\hspace{1cm}
\varphi=-\frac{1+q}{\cot{\ell_2}+q\cot{\ell_1}},
\end{displaymath}
where $q\equiv s_1/s_2$ is an unknown asymmetry parameter. The
135-km\,s$^{-1}$ arm indeed involves higher forbidden velocities than
the 3-kpc arm which, according to our models, means that the former arm
is closer to the Galactic centre than the latter, i.e. $q>1$.
The CO observations suggest $\ell_1=14.5^{\circ}$ and
$\ell_2=-4.5^{\circ}$, yielding the constraints shown in Fig. 1d.
The length $s_1$ increases for smaller inclination angles.
If $\varphi\leq 30^{\circ}$, then $s_1\geq 2.9$~kpc, $q\leq 2$ and,
since $s_1<a$ and $R_{\rm CR}\approx 1.2a$ according to numerical
simulations and early type galaxy analyses, where $a$ is the semi
major axis of the bar, $R_{\rm CR}\geq 3.5$~kpc (for
$R_{\circ}=8$~kpc). Our simulations are presented in details in
Fux~(1998) and some mpeg movies, including live $\ell-V$ diagrams,
are available at http://obswww.unige.ch/\,\~{}fux.
\begin{figure}
\psfig{file=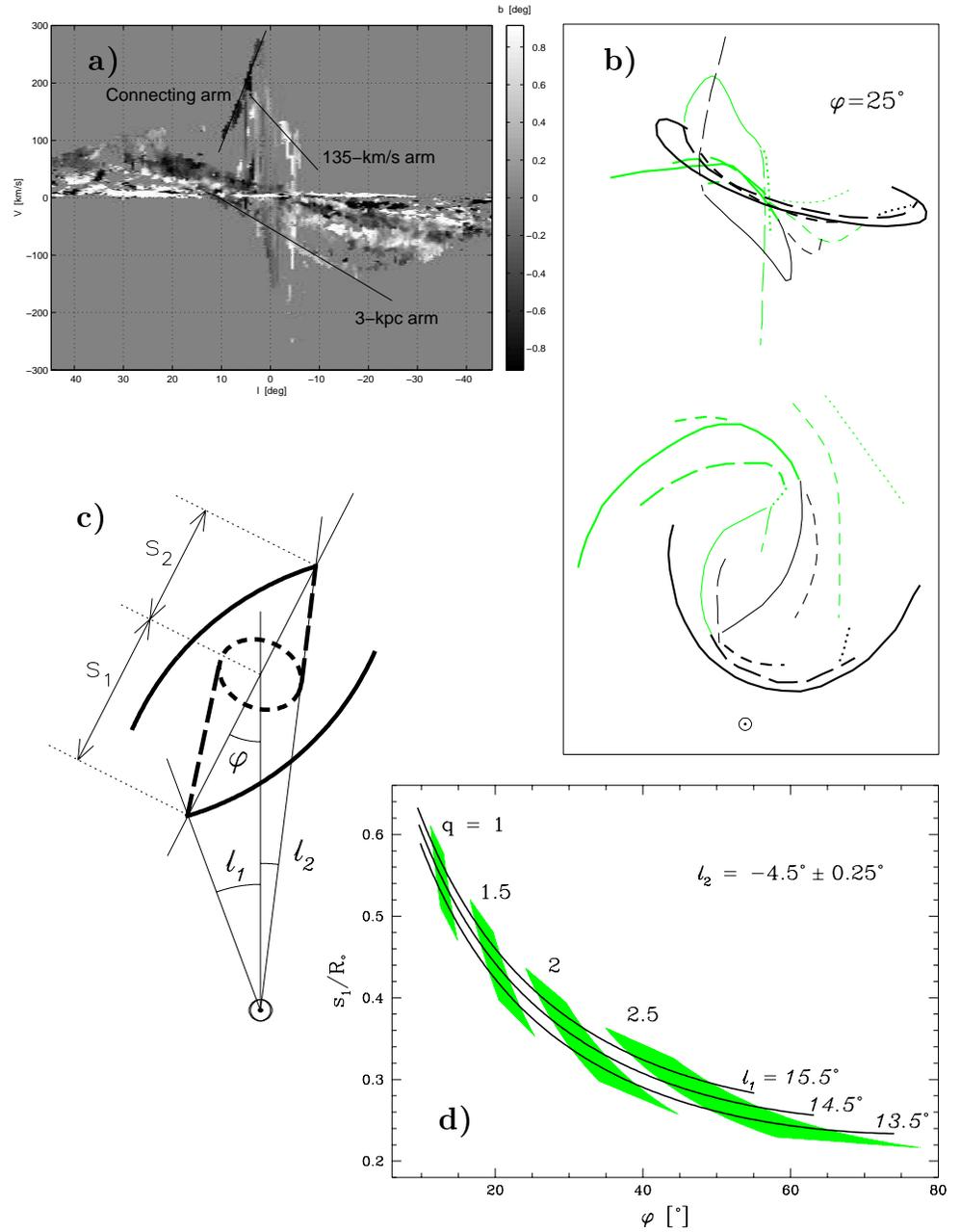,width=13cm}
\caption{a) Observed CO $\ell-V$ diagram (Dame et al. 1987), with the
         grey scale indicating the mean latitude of emission.
         b) Face-on spiral structure of an SPH model (bottom) and
         $\ell-V$ traces of the associated arms (top). c) Parameters
         defining the position of the intersections of the lateral
         arms with the dustlane shocks. The location of the Sun is
         indicated by the $\odot$ symbol. d) Derived constraints on
         these parameters, with $q=s_1/s_2$ and where $R_{\circ}$ is
         the galactocentric distance of the Sun.}
\end{figure}

\end{document}